\documentclass{article}

\usepackage{arxiv}

\usepackage[utf8]{inputenc} 
\usepackage[T1]{fontenc}    
\usepackage{url}            
\usepackage{booktabs}       
\usepackage{amsfonts}       
\usepackage{nicefrac}       
\usepackage{microtype}      
\usepackage{lipsum}		
\usepackage{doi}

\usepackage{color,array,amsthm}
\usepackage{graphicx}
\usepackage{tabularx}
\usepackage{float}
\usepackage{amsmath,amsthm}
\usepackage{epstopdf}
\usepackage{lineno,hyperref}

\newcommand{\angvel}{\boldsymbol\omega}
\newcommand{\linvel}{V}
\newcommand{\genvel}{\boldsymbol{v}}

\newtheorem{asm}{Assumption}
\newtheorem{rem}{Remark}

\newtheorem{myproof}{Proof}

\title{Data-Assisted Control - A Framework Development by Exploiting NASA GTM Platform}

\author{ \href{https://orcid.org/0000-0003-3344-5068}{\includegraphics[scale=0.06]{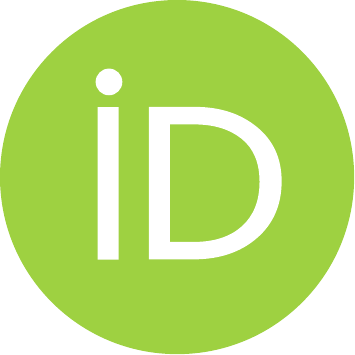}\hspace{1mm}Mostafa Eslami} \\
	Department of Aerospace Engineering\\
	Sharif University of Technology\\
	Azadi Avenue, Tehran \\
	\texttt{eslami.mostafa@ae.sharif.edu} \\
	\And
	\href{https://orcid.org/0000-0000-0000-0000}{\includegraphics[scale=0.06]{orcid.pdf}\hspace{1mm}
 Afshin Banazadeh} \\
	Department of Aerospace Engineering\\
	Sharif University of Technology\\
	Azadi Avenue, Tehran \\
	\texttt{banazadeh@sharif.edu} \\
}

\renewcommand{\shorttitle}{\textit{arXiv} Template}

\hypersetup{
pdftitle={Data-Assisted Control - A Framework Development by Exploiting NASA GTM Platform},
pdfsubject={Systems and Control},
pdfauthor={Mostafa Eslami, Afshin Banazadeh},
pdfkeywords={Data-Assisted Control, GTM},
}

\begin{document}
\maketitle

\begin{abstract}
 Today's focus on expanding the capabilities of control systems, resulting from the abundance of data and computational resources, requires data-based alternatives over model-based ones. These alternatives may become the sole tool for analysis and synthesis. Nevertheless, mathematical models are available to some extent, especially for air and space vehicles. Hypothetically, data assistance would be the approach to meet the requirements in collaboration with the model. In this paper, a framework of Data-Assisted Control (DAC) for aerospace vehicles is proposed. NASA Generic Transport Model (GTM) is the platform for the study and the data supports the model-based controller in extending performance over a damage event. The framework requires real-time decisions to override the control law with the information obtained from the data, while the model-based controller does not show regular performance. The closed-loop system is shown to be stable in the transition phase between the data and the model. The fixed dynamic parameters are estimated using the Dual Unscented Kalman Filter (DUKF) and the evolution of the generalized force moments is estimated using the Koopman estimator. Simulations have shown that the purely model-based robust control leads to degradation of the closed-loop performance in case of damage, suggesting the need for data assistance.
\end{abstract}


\section{INTRODUCTION}
The diversity and volume of the available data, the significant increase in real-time calculation capacity, quality of living in the cyber-physical systems era, human ambitions in interplanetary probing and exploring, and real-time decision-makings in autonomy all have promoted utilizing of the data in the development of science and engineering more and more, the fourth-paradigm \cite{tolle2011fourth}. Before the evolution of model-based analysis and synthesis, data was the first aid for scientists and engineers. Using data in control is not new. One of the first data-driven control design methods in 1942 is Ziegler-Nichols signal-based tuning for PID controllers \cite{ziegler1942optimum}. With progress in physics and our understanding of physical systems, mathematical modeling became the only aid in the development of theories and particularly in control engineering. In control engineering, learning from data is a new paradigm \cite{de2019formulas}. All theories and control methods in which the controllers are derived from the data without explicit use or knowledge of the mathematical model, named Data-Driven Control (DDC). A legitimate DDC is considered to have stability, convergence, and robustness with rigorous mathematical and specified assumptions. The Willems et al.'s Fundamental Lemma (WFL) \cite{willems2005note}, Data Informatively \cite{gevers2009identification,markovsky2007linear}, Behavioral Approach \cite{maupong2017data}, Interconnection Paradigm \cite{willems1997interconnections,trentelman1996control}, Dissipativity \cite{willems1972dissipative,maupong2017lyapunov,berberich2020trajectory}, and data-driven Koopman Operator \cite{williams2015data,budivsic2012applied}, all are provided different frameworks for developing DDC. In the same fashion, Model-Based Control (MBC) explicitly uses the mathematical model. Data-driven control systems may have better performance compared to the MBC whenever the following conditions are the case: a) mathematical model is not available, b) uncertainties and a wide range of structural/nonstructural combinations raise the lack of a comprehensive and integrated model, c) modeling is very hard or impossible such that model-based control won't work well, and d) mathematical model is rather complicated to be used for control system design.

Considering as a fact, always for highly uncertain dynamics, the model-based paradigm has limits in particular for unpredicted operating conditions and situations. These unmodeled dynamics yields unreliable model-based control even with adaptation \cite{anderson2005failures,anderson2007historical,anderson2008challenges}. Otherwise, any increase in the order and complexity of a model may increase the effort and tries for control system design \cite{hou2013model}. The Rohrs' counterexample is an alarm for having too much reliance on adaptation due to unpredictable nonlinear behaviors due to assumptions in model-based controls \cite{rohrs1982robustness,rohrs1985robustness}. Moreover, any dynamical system may show some degree of uncertainties with the increase in operating condition ranges that are not possible to be modeled or identified with high fidelity.

Since data-based theories don't use implicitly any dynamical models and only rely on in-out data of the system, may resolve the requirements for enhanced requirements.
As a fact, the model-based paradigm will work optimally for a certain region with some degree of known uncertain boundaries with very vigorous theories and proven results in abundant practices. However, today's progressive increase in computational capacity has introduced new visions for the evolution of control systems with the assistance of data to extend model-based capabilities. For example the advantage of the data for the enhancement of the model-based control system performance in an industrial engine is evident \cite{eslami2020control}.

Looking at breakthroughs in control engineering, the direct adaptive control \cite{aastrom2013adaptive,idan2002hierarchical} and neural-network-based control systems \cite{werbos1989neural,kim1997nonlinear} are in the same paradigm of the data-driven control systems. After overwhelming use of neural networks in control systems design, the data-driven controllers
were introduced and are progressively developing \cite{hou2013model,bazanella2011data,de2019formulas}. In recent decades, using DDC methods adapted from a similar MBC had great intention among control system scholars. For example PID like DDC \cite{keel2008controller,fliess2013model}, model-reference control and output tracking \cite{campestrini2017data,novara2016data}, predictive controls \cite{salvador2018data}, robust controls \cite{dai2018moments}, Fault-Tolerant Controls (FTC) \cite{yin2013real,emami2020fault}, reinforcement learning \cite{bradtke1993reinforcement}, optimal control \cite{pang2018data,mukherjee2018model,da2018data,baggio2019data} are developed under data-based paradigm.

In aerial applications, the dynamical model with high fidelity is developed \cite{roskam1995airplane}. With a correct understanding of these dynamics, the exact and uncertain parts can be fully distinguished. The variety of operating points in these systems may raise a high degree of uncertainty because all analysis and synthesis methods use simplification rules to express them in a closed form. For example, atmospheric disturbances and turbulence \cite{bieniawski2008micro}, flying in an unknown stochastic and heterogeneous atmosphere or complex environment \cite{ol2008flight}, flights reconfiguration \cite{bacon2000reconfigurable,bai2021low,li2020hybrid}, foreign object damage to the flight's body and control surfaces \cite{liu2010modeling,chowdhary2013guidance}, high-frequency dynamics in control surfaces \cite{chowdhary2013guidance}, increase in bandwidth \cite{hovakimyan2001adaptive}, robustness against faults and uncertainties that are unbounded or impossible to recognize the bound \cite{kamalapurkar2018reinforcement}, envelope protection \cite{yavrucuk2009envelope}, control in case of low speed in communication and low bandwidth of communication which needs real-time decision-makings \cite{badger2019distributed}, and eventually the survivability \cite{jourdan2010enhancing}, all may not be possible to be defined with general deterministic terms in accordance with model-based dynamic formulates.

Moreover, the flights are used to have high reliability with the probability of catastrophic failure less than $10^{-9}$ per hour according to APR4761. For decades, the designers have gained this reliability with hardware redundancy in controllers, and control surfaces, using multi actuators, sensors, and flight computers \cite{yeh1996triple,yeh1998design}. Yet, in 2017, after a series of studies in NASA Langley Research Center \cite{belcastro2017aircraft}, the Loss-of-Control (LOC) \cite{belcastro2016aircraft}, was recognized as the biggest threat to aviation accidents, especially for big commercial fixed wings airplanes. The LOC had the most portion of the accidents across all classes of flights, missions, and flight phases \cite{reveley2010causal,airplanes2016statistical}. One of the significant external reasons for LOC is the damage to the wing and fuselage or control surfaces. Accidents reported flight AAF587 \cite{board2001flight}, A300-B4 \cite{lemaignan2005flying}, AAF232 \cite{flight232}, and BGF1907 \cite{nguyen2008flight} are example of these damages. After studies for analysis and improvement of control systems in face of LOC in \cite{jordan2008nasa,cooper1969use,dydek2010adaptive,bailey2005experimental,johnson2005adaptive}, the adaptive controllers show improvement in flight quality in case of failure and uncertainties \cite{steinberg2001comparison}. Although the adaptive controls have a long and rich history, due lack of simple verification and validation methods, they are not used widely in commercial aerospace applications \cite{nguyen2008flight,jacklin2005development}.

In this paper, a framework is developed in which data extends the flight operating regime of pure MBC in case of the aforementioned uncertainties. The framework utilizes data in the linear evolution of observations thanks to Koopman linearization \cite{koopman1931hamiltonian}. In recent years, the Koopman operator and linearization successfully have been applied in many aerial and space applications such as low-thrust trajectory optimization \cite{hofmann2022advances}, attitude control of spacecraft \cite{chen2020koopman}, studying the motion of
a satellite close to a libration point \cite{servadio2022dynamics,linares2019koopman}, approximating analytical solution for Zontal Harmonics problem \cite{arnas2021approximate}, and decision making \cite{shem2008addressing}. Despite its increasing popularity, the application is very limited due to demanding
accuracy in real-time \cite{servadio2022dynamics}.

True recognition of the part that data can assist, needs the development of a framework from a control engineering point of view. This study is carried out under a specific model-based nonlinear framework with the remedy of a Lyapunov-based nonlinear robust control design, a nonlinear dual estimation with Unscented Kalman Filter (UKF), a data-based Koopman operator over estimated force-moment (pseudo-observations), and a real-time decision-making algorithm with a behavioral approach. The Koopman operator provides powerful data-based linearization in observable space \cite{korda2018linear}, and the nonlinear control benefits from the structure of the nonlinear model capable of stability and performance analysis with a decision factor for the shift between data and model. This compels to use of a nonlinear estimator for parameters and the evolution of forces and moments. We call the exact dynamics with parametric uncertainty, the fixed dynamics. Eventually, the real-time decision-making follows a behavioral approach to optimize an online cost function to promote the data or model credibility, whereas the system behaves like the initial model or undergoes uncertainty or failure. This framework is named Data-Assisted Control (DAC).

DAC has full advantage of the MBC with stability guaranty, yet the data would extend its performance in any condition that flights don't demonstrate the typical behavior. When data interferes, the stability is ascertained with specific assumptions. Of course, when dynamics are certain the data would not interfere. The generic 5.5\% scaled transport class vehicle known as the Generic Transport Model (GTM) by NASA is the platform for this study \cite{jordan2004development,bacon2007general,hueschen2011development}. The GTM includes all subsystems required for experimental flight control algorithm implementation and evaluation \cite{hassan2020design,sun2021command,emami2018multimodel}. The model encompasses a series of defined failures and can be used as a benchmark for FTC systems \cite{nguyen2008flight,liu2010modeling,chowdhary2013guidance}.

The DAC framework is introduced in details in section \ref{sec:DAC}, and required five steps for its development for GTM is provided from sub-sections \ref{sec:GTM} to \ref{sec:decision}. In the end, the framework is applied to the GTM with a series of simulations in section \ref{sec:simulations} to demonstrate the extended performance of DAC compared to the pure MBC.

\section*{NOMENCLATURE}
\begin{table}[H]
\begin{tabular}{|m{5em} m{40em}|}
\hline\hline
$\angvel$  & [rad/sec] angular velocity vector with $p$-roll rate, $q$-pitch rate and $r$-yaw rate arguments in body coordinate \\\cline{2-2}
$\rho$ &  [ft]  center of mass displacement \\\cline{2-2}
$\linvel$ & [ft/sec] linear velocity vector with $u$, $v$ and $w$ arguments in body coordinate\\\cline{2-2}
$\genvel$ & generalized body coordinate velocity vector - $[\linvel^T,\angvel^T]^T$\\\cline{2-2}
$m$ & [lbs] aircraft mass \\\cline{2-2}
$I_M$ & [slug.ft$^2$] aircraft moment of inertia matrix with $I_{ab}$ entries for axis $a$ and $b$\\\cline{2-2}
$I$ & identity matrix with proper dimension\\\cline{2-2}
$g$   & [ft/sec$^2$] gravity constant \\\cline{2-2}
$\eta_1$ & position vector with $X$, $Y$ and $Z$ arguments in earth coordinate\\\cline{2-2}
$\eta_2$ & orientation vector (Euler angles) with $\Phi$, $\Theta$ and $\Psi$ arguments in earth coordinate \\\cline{2-2}
$\eta$ & generalized earth-fixed coordinate position and orientation - $[\eta_1^T,\eta_2^T]^T$ \\\cline{2-2}
$\mathrm{p}$ & parameters set\\\cline{2-2}
$W$ & [lbf] gravitational force \\\cline{2-2}
$F$ & [lbf] external force applied to aircraft \\\cline{2-2}
$T$, $T_R$, $T_L$ & [lbf] total thrusts generated by engines, right engine thrust, left engine thrust \\\cline{2-2}
$M$ & [lbf.ft] external moment applied to aircraft \\\cline{2-2}
$\tau$ & generalized force and moment vector\\\cline{2-2}
$L$ & generalized momentum vector consisting of linear and angular momentums\\\cline{2-2}
$\mathcal{M}$ & mass-inertia matrix\\\cline{2-2}
$\mathcal{C}$ & Coriolis and centrifugal matrix\\\cline{2-2}
$\mathcal{G}$ & gravitational force and moments vector\\\cline{2-2}
$B$ & control derivatives in force and moments\\\cline{2-2}
$D$ & damping derivatives in force and moments\\\cline{2-2}
$C$ & dimensionless aerodynamics coefficient\\\cline{2-2}
$E$ & extra order dynamics\\\cline{2-2}
$e_x$ & indication of error associated with the variable $x$\\\cline{2-2}
$\omega_x$ & state transition noise vector with $\omega_\nu$, $\omega_{\tau_a}$ and $\omega_\mathrm{p}$ arguments \\\cline{2-2}
$\omega_y$ & measurement function noise vector \\\\
\end{tabular}
\end{table}
\begin{table}[H]
\begin{tabular}{|m{5em} m{40em}|}
$\Sigma .$   & summation operator - means total forces or moments \\\cline{2-2}
$\mathcal{S}(.)$ & skew-symmetric operator acts on the vector (.) and generates associated skew-symmetric matrix\\\cline{2-2}
diag$(.)$ & a function generating diagonal matrix of vector (.)\\\cline{2-2}
$\|.\|_2$ & norm-2 of vector (.) or Frobenius norm of matrix (.) \\\cline{2-2}
$.^T$	& transpose operator\\\cline{2-2}
$\hat{.}$ & estimated parameter, vector or matrix\\\cline{2-2}
$\tilde{.}$ & difference between the actual and the desired variable\\\cline{2-2}
Subscripts & \\
$g$ & Indication of the center of gravity\\\cline{2-2}
$p$ & Indication of the center of pressure\\\cline{2-2}
$T$ & Belong to the thrusters\\\cline{2-2}
$A$ & Belong to the aerodynamics\\\cline{2-2}
$F$ & Belong to the forces\\\cline{2-2}
$M$ & Belong to the moments\\\cline{2-2}
$\delta$ & Belong to the control surfaces/actuators\\\cline{2-2}
$er$, $el$, $e$ & right elevator, left elevator, and coordinated left-right elevator together\\\cline{2-2}
$ru$ & rudder\\\cline{2-2}
$ar$, $al$, $a$ & right aileron, left aileron, and coordinated left-right aileron together\\\cline{2-2}
$r$ & reference/residual value\\\cline{2-2}
$a$ & under state variable means augmentation\\\cline{2-2}
$d_i$ & indication of GTM damage case $i$\\\cline{2-2}
$d$ & indication of desired value\\\hline\hline
\end{tabular}
\end{table}

\section{DAC FRAMEWORK}\label{sec:DAC}

\begin{figure}[ht]
  \centering
  \includegraphics[width=\columnwidth]{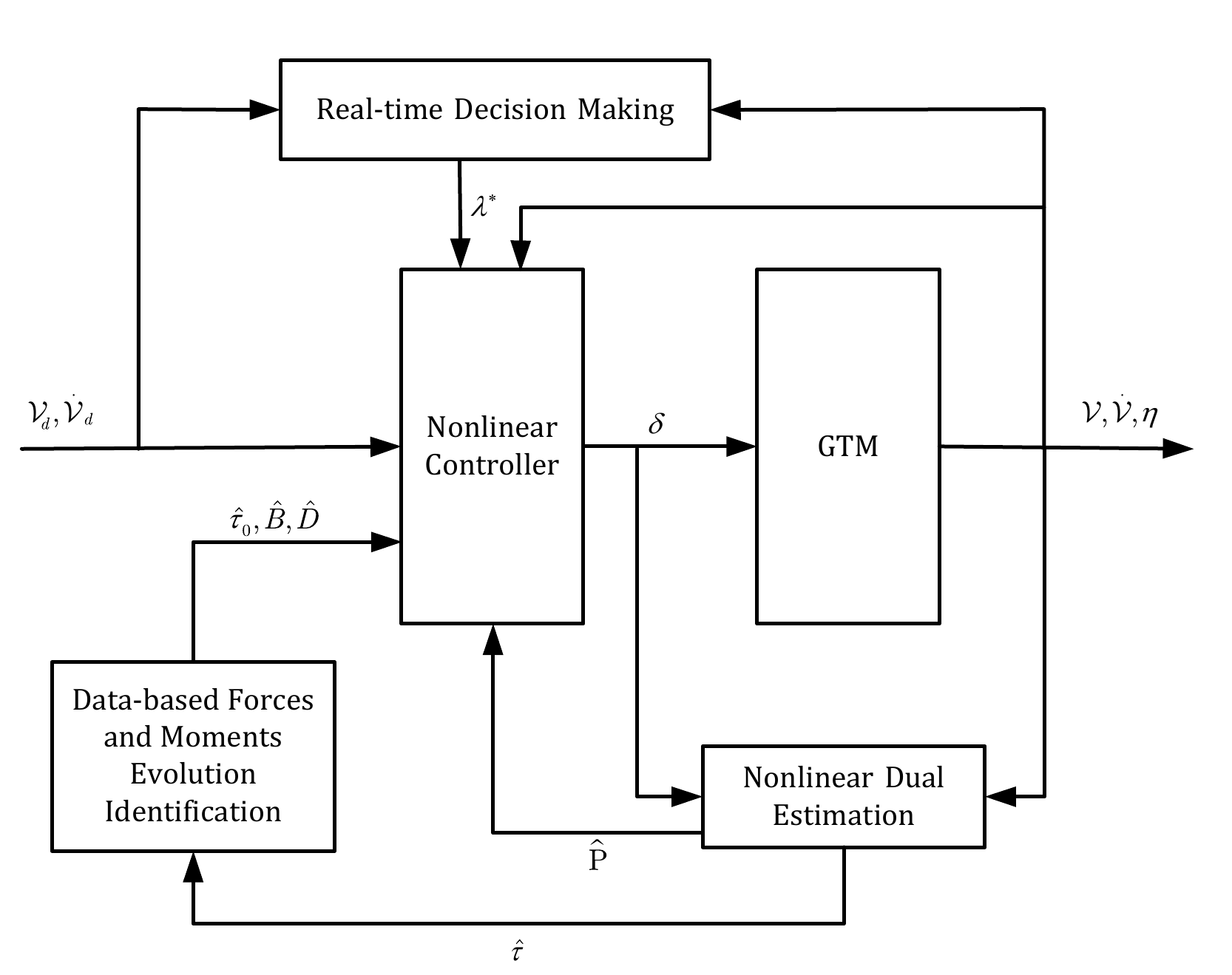}
  \caption{Data Assisted Control (DAC) framework for GTM }\label{fig:DAC_nonlinear_GTM}
\end{figure}

The general framework of the DAC is depicted in Figure \ref{fig:DAC_nonlinear_GTM}. In this scheme, considering the fixed structure for $\mathcal{M}$ and $\mathcal{C}$ matrices, parameter estimation is used for the estimation of mass, the moment of inertia, and their products. In this case, any effect on the force and moment derivatives is reflected in applied aerodynamics and thrust forces and moments, $\tau$. One can call this parameter pseudo-observation. Because first using the measurement of states a nonlinear estimate is exploited to use the fixed dynamics and then the estimated force-moment vector is used as observation. Due to the structure of the developed model, the fixed dynamics are strictly equivalent to the observations, but $\tau$ is not measured in practice. Accordingly, the dual estimation will help to have correct estimates of the generalized forces and moments evolution by time without the need for a direct measurement. Simply, the fixed dynamics with correct parametric and state estimates must provide correct trends of external forces and moments. Eventually, the linear evolution of the observation of the generalized forces and moments will be identified and fed to the control law to adapt the closed-loop dynamics. According to the force expansion rule by \cite{hopkin1968scheme,zipfel2000modeling}, the generalized forces and moments are characterized by linear dependency on aerodynamic derivatives up to some unknown vanishing orders. Therefore, the regressor will have significant void entries which promote using data-based methods for identification over observations. The real-time decision-making algorithm always tracks the performance of the closed-loop system and with slight changes in the decision factor tests interference of the identified regressions by data. If the external behavior of the aircraft demonstrates improvement in the performance, the decision factor will promote the data assistance more and more, otherwise, it will return to the MBC. The DAC framework can be decomposed into modular steps as follows:
\begin{description}
  \item[(A) ] rewriting the nonlinear model suite for DAC framework and performing dynamics decoupling for observations and fixed dynamics,
  \item[(B) ] nonlinear control design for the full envelope, using fixed dynamics,
  \item[(C) ] parameter estimations including mass, the moment of inertia, the center of gravity, and pseudo observation estimation considering the fixed-dynamics exploiting a dual estimation approach,
  \item[(D) ] data assistance for identification of the aerodynamic and control derivatives over pseudo observations,
  \item[(E) ] behavioral decision-making for optimizing decision factor of altering control law according to the data-assisted model in a stable way.
\end{description}

This framework is elaborated through a series of simulations and analysis in a pragmatic way to provide seemly data assistance with stability and performance guarantee.

\subsection{GTM Nonlinear Model Suitable for DAC}\label{sec:GTM}
The set of flight dynamics equations for GTM is developed considering the change in the body's center of mass. Such equations can be used when the body loses a portion of its mass and it is desired to track the motion of the body's previous center of mass/reference frame \cite{bacon2007general}. The velocity components are subscripted with any arbitrary point in body-frame to distinguish them from the usual velocity components considered at the center of mass. Of course these components are the velocity of the center of mass when this point is the center of mass, i,e. $\rho = 0$. The force of gravity and the equation of motion is written as follows \cite{bacon2007general,guo2011multivariable},,
\begin{align}
  W = \left[
             \begin{array}{c}
               -mg\sin \Theta \\
               mg\cos \Theta \sin \Phi \\
               mg \cos \Theta \cos \Phi \\
             \end{array}
           \right]
\end{align}
\begin{align}
  &\Sigma F + W= m\left(\dot{\linvel}+\mathcal{S}(\angvel) \linvel\right)-m\mathcal{S}(\rho)\dot{\angvel} + m\mathcal{S}(\angvel)(\mathcal{S}(\angvel)\rho)\nonumber\\
  &\Sigma M + \mathcal{S}(\rho) W = I_M\dot{\angvel}+\mathcal{S}(\angvel) (I_M \angvel) + m \mathcal{S}(\rho) \dot{\linvel}+\nonumber\\
  &m \mathcal{S}(\angvel)(\mathcal{S}(\rho) \linvel)+m \mathcal{S}(\linvel)(\mathcal{S}(\angvel)\rho)
\end{align}




The derived equation of motion can be concisely described in a matrix form as below,
\begin{align}\label{equ:eom_closed}
  \mathcal{M}\dot{\genvel}+\mathcal{C}(\genvel)\genvel + \mathcal{G}(\eta) = \tau
\end{align}

where $\mathcal{M}$ is the mass-inertia matrix, $\mathcal{C}$ is Coriolis and centrifugal matrix, and $\mathcal{G}$ is gravitational force and moment acting on the flight as,
\begin{align}
  &\mathcal{M} = \left[\begin{array}{cc}
               mI & -m\mathcal{S}(\rho)\\
               m\mathcal{S}(\rho) & I_M\\
             \end{array}
           \right],\nonumber\\ &\mathcal{C}(\genvel)=\left[\begin{array}{cc}
               m\mathcal{S}(\angvel) & -m\mathcal{S}(\mathcal{S}(\angvel)\rho)\\
               -m\mathcal{S}(\mathcal{S}(\angvel)\rho) & -\mathcal{S}(I_M\angvel)+m\mathcal{S}(\mathcal{S}(\linvel)\rho)\\
             \end{array}
           \right]\;\&\nonumber\\
&\mathcal{G}(\eta) = \left[\begin{array}{c}
               -W\\
               -\mathcal{S}(\rho)W\\
             \end{array}
           \right].
\end{align}

\begin{rem}\label{rem:mdot_2c}
  Assuming small change rate in mass and moment of inertia, $\mathcal{M}$ and $\mathcal{C}$ are rearranged such that $\dot{\mathcal{M}}-2\mathcal{C}(\genvel)$ is skew-symmetric.
\end{rem}

The following decomposition can be made over GTM forces and moments produced by engines ($F_T$ and $M_T$), the aerodynamics of wings and base airframe ($F_A$ and $M_A$), and control surfaces ($F_\delta$ and $M_\delta$),
\begin{align}
  &\Sigma F = F_T + F_A + F_\delta, \text{ and}\nonumber\\
  &\Sigma M = M_T + M_A + M_\delta + \mathcal{S}(l-\rho)(F_A+F_\delta)
\end{align}

Where $l=C_p-C_g$ is difference between center of pressure and gravity (for GTM $l = [-0.0276, 0.0118, 0.036]^T$ [ft]). The forces and moments of the aerodynamic base frame and control surfaces are traditionally defined by dimensionless aerodynamic coefficients, $C_F$, and $C_M$. Each coefficient is decomposed into state variables with a set of multipliers. The angle of attack $\alpha$ and side-slip angle $\beta$ for aerodynamics of base frame and control surface angle purely define the flight aerodynamic attitude with respect to surrounding air and direction. Hence, it is common to define the multipliers for these two angles, not the state variables. The relation between aerodynamic coefficients and force and moments in reference flight is as follows,
\begin{align}
  &F_A = C_F\bar{q}_rS_r,\quad F_\delta=C_{\delta_F}\bar{q}_rS_r,\nonumber\\
  &M_A = \text{diag}(b,\bar{c},b)\times C_M\bar{q}_rS_r, \nonumber\\
  &M_\delta = \text{diag}(b,\bar{c},b)\times C_{\delta_M}\bar{q}_rS_r.
\end{align}

Where,
\begin{align}
  & C_A = \left[\begin{array}{c}
               C_{F}\\
               C_{M}\\
             \end{array}
           \right] = C_0+C_\alpha\alpha+C_\beta\beta\quad \text{and}\nonumber\\
  & C_{\delta_A} = \left[\begin{array}{c}
               C_{\delta_F}\\
               C_{\delta_M}\\
             \end{array}
           \right] = \left[\begin{array}{ccc}
               C_{\delta_{ru}}\delta_{ru} & C_{\delta_{a}}\delta_{a} & C_{\delta_{e}}\delta_{e}\\
             \end{array}
           \right].
\end{align}

\begin{asm}
  The parameters are extracted for reference flight of GTM model in $800$ [ft] altitude and TAS of 75 [knots] without the wind,  $\bar{q}_rS_r=105.65$ [lbf], $b=6.8488$ [ft], and $\bar{c}=0.9153$ [ft].
\end{asm}

\begin{asm}
In GTM the aerodynamic coefficients and control derivatives are generated via a look-up table for $-5\leq\alpha\leq+85$ DEG and $-45\leq\beta\leq+45$ DEG. In the rest, we assume that flight is in steady symmetry condition near $\alpha=4$ DEG and $\beta=0$ DEG.
\end{asm}


\begin{asm}
It is assumed flaps, stabilizers, spoilers, brake, and landing gear don't participate in flight. Their trim value in simulated reference flight is zero. It is also assumed engines are symmetrical so they generate the same thrust value and we have a single $\delta_T$ as a thrust control input. The upper and lower rudder work together, and single control input $\delta_{ru}$ is considered for the rudder. For aileron and elevator, the inner and outer boards and the left and right surfaces are merged to accept only one control input as $\delta_{a}$ for aileron and $\delta_{e}$ for elevator. The prime motive behind this control input configuration is the possibility of damage simulations in GTM standard damage models. In summary vector of control input is as follows,
\begin{align}
  \delta = \left[\begin{array}{c}
               \delta_T\\
               \delta_{ru}\\
               \delta_{a}\\
               \delta_{e}\\
             \end{array}
           \right]
\end{align}
\end{asm}

Considering the assumptions in the derived model, the simplified mathematical representation of the GTM in simulations will have residual forces and moments in the dynamics. Let's denote the residual force and moments in reference flight by $\tau_r$. The final form of the nonlinear model will be summarized as,
\begin{align}\label{equ:eom_closed2}
  \mathcal{M}\dot{\genvel}+\mathcal{C}(\genvel)\genvel + \mathcal{G}(\eta) = \tau + \tau_r.
\end{align}

\subsection{Nonlinear Robust Control Design - a Velocity Regulator}\label{sec:control_law}
Equation \eqref{equ:eom_closed2} ensembles dynamics in a ready form for developing nonlinear controller. The objective of the control system design in case of failure is presumed to be regulation of the velocities in body coordinate, as a pilot would act.  Accordingly, the design of a nonlinear velocity regulator has been followed next. Let's define an energy Lyapunov candidate function as follows,
\begin{align}\label{eq:lyapunov}
  \mathcal{V} = \dfrac{1}{2}\tilde{\genvel}^T\mathcal{M}\tilde{\genvel}
\end{align}

Where $\tilde{\genvel}=\genvel-\genvel_d$. Taking derivative from Lyapunov function and using Remark \ref{rem:mdot_2c} yields,
\begin{align}
  \dot{\mathcal{V}} &= \tilde{\genvel}^T\mathcal{M}\dot{\tilde{\genvel}}+\dfrac{1}{2}\tilde{\genvel}^T\dot{\mathcal{M}}\tilde{\genvel} = \tilde{\genvel}^T(\mathcal{M}\dot{\genvel}-\mathcal{M}\dot{\genvel}_d+\mathcal{C}(\genvel)\tilde{\genvel})\nonumber\\
  &=\tilde{\genvel}^T(\tau+\tau_r-\mathcal{G}(\eta)-\mathcal{C}(\genvel)\genvel-\mathcal{M}\dot{\genvel}_d+\mathcal{C}(\genvel)\tilde{\genvel})
\end{align}

Assume following control law has been selected,
\begin{align}
  \tau = \mathcal{G}(\eta)+\mathcal{C}(\genvel)\genvel_d + \mathcal{M}\dot{\genvel}_d -\Gamma\tilde{\genvel}-\tau_r
\end{align}

Where $\Gamma = \text{diag}(\gamma_1,\gamma_2,...,\gamma_6)$ and $\gamma_i>0$. Then,
\begin{align}
  \dot{\mathcal{V}} = -\tilde{\genvel}^T\Gamma\tilde{\genvel} \leq -\underline{\lambda}(\Gamma)\|\tilde{\genvel}\|^2
\quad \Rightarrow\nonumber\\
  \mathcal{V}\leq \dfrac{1}{2}\overline{\lambda}(\mathcal{M})\|\tilde{\genvel}\|^2 \Rightarrow \|\tilde{\genvel}\|^2 \geq \dfrac{2\mathcal{V}}{\overline{\lambda}(\mathcal{M})}
\end{align}

Hence,
\begin{align}
  \dot{\mathcal{V}} \leq -2\left(\dfrac{\underline{\lambda}(\Gamma)}{\overline{\lambda}(\mathcal{M})}\right)\mathcal{V} \Rightarrow \mathcal{V}(t)\leq \mathcal{V}(0)\exp^{-2\left(\dfrac{\underline{\lambda}(\Gamma)}{\overline{\lambda}(\mathcal{M})}\right)t}
\end{align}

Where $\mathcal{V}(0)\geq 0$. The role of $\underline{\lambda}(\Gamma)$ now is evident. By increasing its value, $\mathcal{V}(t)$ converges to zero exponentially faster. This yields exponential convergence of generalized velocity error towards the origin, with $\Gamma$ as the tuning parameter.

According to GTM, the generalized force and moment vector can be decomposed as follows,
\begin{align}\label{equ:tau_decomposed}
  \tau = \tau_0 + B(\delta)\delta + D(\genvel)\genvel
\end{align}

Where $\tau_0$ is static forces and moments in the trim condition, $B(\delta)$ is a matrix with the variable sign of elements according to the sign of command $\delta$, and $D(\genvel)$ is damping term of the forces and moments linearly dependent to state vector $\genvel$. This yields an update in control law as follows,
\begin{align}
  \tau_c = \mathcal{G}(\eta)+\mathcal{C}(\genvel)\genvel_d + \mathcal{M}\dot{\genvel}_d -\Gamma\tilde{\genvel}-\tau_r - \tau_0 - D(\genvel)\genvel
\end{align}

In which, $\tau_c = B(\delta)\delta$. As a challenge in this control law, the subtracted terms from the control law are open-loop. Yet, the exact evaluation of these values is almost impossible. Hence, additional terms are added to the control law to overcome issues of possible uncertainty and make the control system robust. Considering the $\chi$ as bound of uncertainty, following the control law will make the control system robust \cite{eslami2019robust}.
\begin{align}
  &\tau_c = \mathcal{G}(\eta)+\mathcal{C}(\genvel)\genvel_d + \mathcal{M}\dot{\genvel}_d -\Gamma\tilde{\genvel} -\tau_r- \tau_0 - D(\genvel)\genvel \nonumber\\
  &- \chi\tanh \left(\dfrac{\tilde{\genvel}}{\epsilon}\right)
\end{align}


Eventually, having $\tau_c$ in hand, the calculation of the control command $\delta$ is the last part of the control system computation. Using least square minimization, the optimal control command is calculated as follows,
\begin{align}
  \delta^\ast = (B(\delta)^TB(\delta))^{-1}B(\delta)^T\tau_c
\end{align}

For not damaged aircraft, considering merging of left-right actuators, we have $B(\delta)\in \mathcal{R}^{6\times 4}$ and $\delta\in\mathcal{R}^{4\times 1}$. In case of damage with losing actuators the dimension of the input matrix may decrease.

\begin{rem}
  In case of additional nonlinear term $E$ in $\tau$ as $\tau =\tau_0 + B(\delta)\delta + D(\genvel)\genvel +E$, the term can be decomposed into $E = B_E\delta + H.O.T.$ which are varying by time. The control law is still valid considering new control derivative matrix $B'(\delta)=B(\delta)+B_E$. The high-order terms are subtracted from $\tau_c$.
\end{rem}
\begin{rem}\label{rem:pe}
  In order to ensure persistence of excitation in this regulator, signals consisting of sinusoids of varying frequencies and randoms are added to the control law \cite{vamvoudakis2010online}.
\end{rem}

\subsection{Damaged Aircraft State and Parameter Estimation}
In this study, the aircraft experiences damage case 1 of GTM. Damage case 1 involves a parametric change in mass, moments of inertia and their products, the center of gravity, and force-moment control derivatives. The change in parameters is defined as follows,
\begin{align}
  &m_{d_1} = m - 0.13 \\\nonumber
  &\rho_{d_1} =  [0.0105, 0, 0.0023]^T\\\nonumber
  &I_{xx_{d_1}} = I_{xx} -0.00346\\\nonumber
  &I_{yy_{d_1}} = I_{yy} -0.06698\\\nonumber
  &I_{zz_{d_1}} = I_{zz} -0.06352\\\nonumber
  &I_{xz_{d_1}} = I_{xz} -0.01409\\\nonumber
  &I_{yz_{d_1}} = I_{yz} +0.00001\\\nonumber
  &I_{xy_{d_1}} = I_{xy} +0.00003\nonumber
\end{align}

In this step, the parameters of the fixed dynamics $\mathrm{p} = [m, I_{xx}, I_{yy}, I_{zz}, I_{xz}, I_{yz}, I_{xy}, \rho]^T$, and generalized forces and moments are estimated with the augmented mathematical model. Considering the DUKF approach \cite{wan1999dual}, the state estimation model can be written as,
\begin{align}
  & \dot{\hat{\genvel}} = \hat{\mathcal{M}}^{-1}\left(\hat{\tau}+\tau_r-\hat{\mathcal{C}}\hat{\genvel}-\hat{\mathcal{G}}\right)+\omega_\nu\\\nonumber
  & \dot{\hat{\tau}}_a = A_{\tau_a}\tau_a + \omega_{\tau_a}
\end{align}
and measurement function can be written as $y = \genvel + \omega_y$.
The augmented force-moment, i.e. $\tau_a\in\mathcal{R}^{18\times 1}$, represents third order Gauss-Markov process with auxiliary variables $\zeta_1$ and $\zeta_2$ as follows \cite{sri1988determination},
\begin{align}
  &\tau_a  = \left[
             \begin{array}{c}
               \tau \\
               \zeta_1 \\
               \zeta_2 \\
             \end{array}
           \right] \quad\&\nonumber\\
  & \left[
             \begin{array}{c}
               \dot{\tau} (i) \\
               \dot{\zeta}_1 (i) \\
               \dot{\zeta}_2 (i) \\
             \end{array}\right]= \left[
                             \begin{array}{ccc}
                               0 & 1 & 0 \\
                               0 & 0 & 1 \\
                               0 & 0 & 0 \\
                             \end{array}
                           \right]\left[
             \begin{array}{c}
               \tau (i) \\
               \zeta_1 (i) \\
               \zeta_2 (i) \\
             \end{array}\right] \nonumber\\+  &\left[
                                                         \begin{array}{c}
                                                           \omega_\tau(i) \\
                                                           \omega_{\zeta_1}(i) \\
                                                            \omega_{\zeta_2}(i) \\
                                                         \end{array}
                                                       \right]\quad \text{for }i=1,\ldots,6
\end{align}
Therefore,
\begin{align}
  A_{\tau_a} = \left[
                 \begin{array}{ccc}
                   \emptyset_6 & I_6 & \emptyset_6 \\
                   \emptyset_6 & \emptyset_6 & I_6 \\
                   \emptyset_6 & \emptyset_6 & \emptyset_6 \\
                 \end{array}
               \right],\quad \&\quad \omega_{\tau_a} = \left[
                                                         \begin{array}{c}
                                                           \omega_\tau \\
                                                           \omega_{\zeta_1} \\
                                                            \omega_{\zeta_2} \\
                                                         \end{array}
                                                       \right]
\end{align}

Where, $\emptyset_6$ is zero matrix with dimension of $\mathcal{R}^{6\times 6}$, and $I_6$ is identity matrix with dimension of $\mathcal{R}^{6\times 6}$. The considered process for evolution of parameters in time is considered as follows,
\begin{align}
    \dot{\hat{\mathrm{p}}} = \omega_{\mathrm{p}}
\end{align}


Due to nonlinearity in parameters which results in the whole process with non-additive noises, augmentation is necessary. We define the covariance matrix of zero-mean Gaussian noises as $\omega_{x_a} \sim\mathcal{N}(0,Q)$ and $\omega_y \sim \mathcal{N}(0,R)$ such that $Q,R\succ 0$ to overcome numerical instabilities.

Having the augmented nonlinear model in hand, the DUKF estimation algorithm is followed. In order to avoid the complexity of DAC parameter tuning, the simplest form of sigma-points generation for unscented transform is considered \cite{simon2006optimal}.

\begin{rem}\label{rem:AUKF}
 The UKF algorithm is optimal when a) the model matches the real system perfectly, b) the entering noise is white (uncorrelated), and c) the covariances of the noise are known exactly.
\end{rem}

The framework implies the model matching and uncorrelated noises; however, adaptation in covariances is essential. Using innovation and residual of the filter following adaptation rule is employed to dynamically update noise covariance matrices \cite{akhlaghi2017adaptive},
\begin{align}
  & Q_k = \alpha Q_{k-1} + (1-\alpha)K_kd_kd_k^TK_k^T,\nonumber\\&\quad \text{innovation at step $k$: }d_k = y_k-h(\hat{x}^-_k)\\\nonumber
  & R_k = \alpha R_{k-1} + (1-\alpha)(\varepsilon_k\varepsilon_k^T+H_kP_k^-H_k^T),\nonumber\\&\quad \text{residual at step $k$: }\varepsilon_k = y_k-h(\hat{x}^+_k)
\end{align}
Where in these equations $\alpha$ is a forget factor, and $H_k$ is jacobian of measurement function, $h(.)$, at step $k$.
In addition to the above conditions, the estimation model itself should be observable. In order to check observability let's collect the state transition and measurement functions as follows:
\begin{align}\label{equ:estimation_model}
 & \dot{\genvel} = f(\genvel,\tau,\mathrm{p})\\\nonumber 
  & \dot{\tau}_a = A_{\tau_a}\tau_a + \omega_{\tau_a}\\\nonumber
 & \dot{\mathrm{p}} = \omega_{\mathrm{p}} ,\quad\text{and}\quad y = \genvel + \omega_y
\end{align}
In this study, using Lie derivatives for deriving analytical implications of the observability was found to be inefficient due to demanding high-order partial differentiation. Moreover, these derivatives will induce computational errors and hence declines the credibility of the result from the computation point of view. As an alternative approach due to working in a regulation scheme the piece-wise linearity of the dynamics is assumed, then the analytical linear system is elaborated to check the observability in each step. Considering estimation model \eqref{equ:estimation_model}, the linear system observability matrix couple $(A,C)$ would be,
\begin{align}
    A = \left[
                 \begin{array}{ccc}
                   \dfrac{\partial f}{\partial\genvel} & \dfrac{\partial f}{\partial\tau_a} & \dfrac{\partial f}{\partial\mathrm{p}} \\
                   \emptyset & A_{\tau_a} & \emptyset \\
                   \emptyset & \emptyset & \emptyset \\
                 \end{array}
               \right]\quad\text{and}\quad C =  \left[
                 \begin{array}{ccc}
                   I_6 & \emptyset & \emptyset\\
                 \end{array}
               \right].
\end{align}
It is evident that if $\dfrac{\partial f}{\partial\mathrm{p}} =0$ then model is not observable. Derivatives of $f$ are derived as follows:
\begin{align}
    &\dfrac{\partial f}{\partial\genvel} = -\mathcal{M}^{-1}(\mathrm{p})\underline{\dfrac{\partial }{\partial\genvel}\left(\mathcal{C(\genvel)\genvel}\right)},\quad \dfrac{\partial f}{\partial\tau_a} = \left[
                 \begin{array}{cc}
                   \mathcal{M}^{-1}(\mathrm{p}) & \emptyset\\
                 \end{array}
               \right],\nonumber\\
    &\quad\text{and}\quad
    \dfrac{\partial f}{\partial\mathrm{p}} =\underline{ -\dfrac{\partial}{\partial\mathrm{p}}\left(\mathcal{M}^{-1}(\mathrm{p})\left(\mathcal{C}(\mathrm{p})\genvel+\mathcal{G}(\mathrm{p})\right)\right)}.
\end{align}
The underlined terms are calculated with symbolic calculations, then the observability matrix singular values are checked per iteration. The observability matrix is defined as,
\begin{align}
\mathcal{O} = \left[
                 \begin{array}{c}
                   C  \\
                   CA  \\
                   \vdots\\
                   CA^{n-1}  \\
                 \end{array}
               \right]
\end{align}

\subsection{ Data Assistance}
Koopman demonstrated that it is possible to represent a nonlinear dynamical system in terms of an infinite-dimensional linear operator acting on a Hilbert space of measurement functions of the state of the system \cite{koopman1931hamiltonian}.  Koopman operator is linear, and its spectral decomposition completely characterizes the behavior of a nonlinear system \cite{schmid2010dynamic,brunton2019data}. Using the Koopman operator nonlinear dynamics become completely linear in eigenfunction coordinates, called lifted space.

Provided estimates for generalized force-moment vector and parameters, in this step the control derivatives, damping derivatives, and static force-moment terms are recognized in pseudo observations via measured data. The approach of using Koopman operator in synthesis and design step of the controller is growing rapidly.  In particular, the Koopman estimator is used to provide the linear estimate, and the results are compared with Recursive Least Square (RLS) \cite{aastrom2013adaptive}. The RLS is chosen as an analog due to being an optimal deterministic estimator for linear fitting in adaptive rules. The torque and moments are regularly treated as linearly varying to actuator command, linear and angular velocities with a static term. As a general comparison, the Koopman represents a batch data identification method and the latter updates the estimates according to just previous observations.


From the developed framework the $i$-th observations and measurements in the simplest from are decomposed as follows,
\begin{align}
\hat{\tau}_i = \hat{\tau}_0+\hat{B}(\delta_i)\delta_i+\hat{D}(\genvel_i)\genvel_i
\end{align}
and stack of $m$ previous sampled pseudo observations and measurements are collected as follows,
\begin{align}\label{eq:observation}
  &\hat{\mathcal{T}}=\begin{bmatrix}
   \hat{\tau}_1 & \ldots & \hat{\tau}_m
  \end{bmatrix}\in\mathcal{R}^{6\times m} \quad\&\nonumber\\
  &\mathcal{Y} = \begin{bmatrix}
    \delta_1 & \ldots & \delta_m\\
    \genvel_1 & \ldots & \genvel_m\\
    1 & \ldots & 1\\
  \end{bmatrix}\in\mathcal{R}^{11\times m}.
\end{align}
Accordingly, using Koopman the linear evolution of the observations with respect to the measured variables can be obtained as follows,
\begin{align}
  \hat{\mathcal{P}} = \hat{\mathcal{T}}\mathcal{Y}^T(\mathcal{Y}\mathcal{Y}^T)^{-1}
\end{align}

Where,
\begin{align}
    \hat{\mathcal{P}} =   \begin{bmatrix}
    \hat{B} & \hat{D} & \hat{\tau}_0\\
    \end{bmatrix}\in\mathcal{R}^{6\times 11}.
\end{align}

\begin{rem}
  In practice, the equation \eqref{eq:observation} can be described in a more complex form depending on some orders of the measurement derivatives \cite{zipfel2000modeling}. In this case, a more complicated data-driven algorithm can be used for linear estimation such as SINDy \cite{brunton2016sparse}. Without loss of generality, in this paper, the simplest form is assumed. In order to compare Koopman estimator performance versus RLS, in section \ref{sec:simulations} additional nonlinear terms are appended to the model. The additional terms are put into extra order dynamics vector, $E$, and includes $\sin(10r)\delta_r$.
\end{rem}

\begin{rem}
  Equation \eqref{eq:observation} gives dynamics of the generalized momentum, $\tau = \mathrm{d}L/\mathrm{d}t$. Therefore, the evolution of observations describes the dynamical changes in momentum directly. This can be used to design a Koopman data-driven control system to control momentum directly \cite{korda2018linear}. In this paper, data only assists in control system design, and this approach is not followed. However, as a third option in the decision-making, the full data-driven control like this approach would be the case for the next studies.
\end{rem}


\subsection{Real-time Decision-Making and Decision Factor Optimization}\label{sec:decision}
A decision factor $\lambda$ determines the active role of data assistance. From equation \eqref{equ:eom_closed2} and \eqref{equ:tau_decomposed} the integrated open-loop dynamics is as follows,
\begin{align}
    \mathcal{M}\dot{\genvel}+\mathcal{C}(\genvel)\genvel + \mathcal{G}(\eta) - \tau_r = \tau  = \tau_0 + B(\delta)\delta + D(\genvel)\genvel
\end{align}

This equation after estimations and exploiting data becomes,
\begin{align}
    \hat{\mathcal{M}}\dot{\genvel}+\hat{\mathcal{C}}(\genvel)\genvel + \hat{\mathcal{G}}(\eta) - \tau_r = \hat{\tau} = \hat{\tau}_0 + \hat{B}(\delta)\delta + \hat{D}(\genvel)\genvel
\end{align}

Using decision factor $\lambda$, the coupled estimated and initial/uncertain dynamics will be as follows,
\begin{align}\label{eq:coupled}
    &\mathcal{M}(\lambda)\dot{\genvel}+\mathcal{C}(\lambda)\genvel + \mathcal{G}(\lambda)  - \tau_r = \tau(\lambda) = \tau_0 (\lambda) + B(\lambda)\delta \nonumber\\&+ D(\lambda)\genvel
\end{align}

Where,
\begin{align}\label{eq:lambda_par}\centering
 & \mathcal{M}(\lambda) = (1-\lambda)\mathcal{M}+\lambda\hat{\mathcal{M}}\\
 & \mathcal{C}(\lambda) = (1-\lambda)\mathcal{C}+\lambda\hat{\mathcal{C}}\nonumber\\
 & \mathcal{G}(\lambda) = (1-\lambda)\mathcal{G}+\lambda\hat{\mathcal{G}}\nonumber\\
 & \tau(\lambda) = (1-\lambda)\tau+\lambda\hat{\tau}\nonumber\\
 & \tau_0(\lambda) = (1-\lambda)\tau_0+\lambda\hat{\tau}_0\nonumber\\
 & B(\lambda) = (1-\lambda)B+\lambda\hat{B}\nonumber\\
 & D(\lambda) = (1-\lambda)D+\lambda\hat{D}\nonumber
\end{align}

The decision factor is a convex combination of the initial/uncertain dynamics and the estimated one. As $\lambda\rightarrow 0$ control apportioned to MBC, otherwise, $\lambda\rightarrow 1$ suggests uncertainty has appeared and DAC would become operative.

For decision-making, the historical behavior of the closed-loop system is turned into a performance cost function with $\lambda$ as optimization variable in real-time. Also, the decision on $\lambda$ can be taken manually by pilot. Assuming time horizon of $m_\lambda$ samples in past period $t_p\leq t\leq t_0$, following cost function is a candidate to select $\lambda^\ast$,
\begin{align}
  J_\lambda = \dfrac{1}{2}\tilde{\genvel}^T(\lambda,t_0)H_\lambda\tilde{\genvel}(\lambda,t_0)+\dfrac{1}{2}\int_{t_p}^{t_0} \tilde{\genvel}^T(\lambda,t)Q_\lambda\tilde{\genvel}(\lambda,t)\mathrm{d}t
\end{align}

subject to dynamics in \eqref{eq:coupled} considering $\lambda = \lambda^{\ast - }$ as the optimum $\lambda$ in the previous window $t_0-2t_p\leq t\leq t_p$. The $H_\lambda$ and $Q_\lambda$ are weightings for terminal and transient cost.

In this study the gradient based steepest decent is used for minimization of $J_\lambda$ \cite{nocedal2006numerical}. In steepest decent the next optimization variable is updated by the following rule,
\begin{align}
  \lambda_{k+1}^\ast = \lambda_{k}^\ast  - \gamma_k \dfrac{\nabla J_\lambda(\lambda_{k}^\ast)}{\|\nabla J_\lambda(\lambda_{k}^\ast)\|}
\end{align}
where $\gamma_k$ is step size gain, and $k$ is the iteration of the optimization. Step size gain in companion with decision horizon determines the rate of change in decision factor.  Legitimate decision-making should have a stable transition period from data to model and vice versa. Next, it is shown under certain conditions, the closed-loop system in DAC is stable.
\begin{myproof}
Considering $\dot{\mathcal{M}}(\lambda)-2\mathcal{C}(\lambda)$ is still skew-symmetric, the developed control law in Section \ref{sec:control_law} is still valid and can be written as,
\begin{align}
  &\tau_c(\lambda^\ast) = \mathcal{G}(\lambda^\ast)+\mathcal{C}(\lambda^\ast)\genvel_d + \mathcal{M}(\lambda^\ast)\dot{\genvel}_d -\Gamma\tilde{\genvel}  - \tau_r \nonumber\\
  &- \tau_0(\lambda^\ast) - D(\lambda^\ast)\genvel - \chi\tanh \left(\dfrac{\tilde{\genvel}}{\epsilon}\right)
\end{align}
Therefore the stability analysis falls into checks on the skew-symmetric property of $\dot{\mathcal{M}}(\lambda)-2\mathcal{C}(\lambda)$. Accordingly using \eqref{eq:lambda_par},
  \begin{align}
    &\dot{\mathcal{M}}(\lambda) - 2\mathcal{C}(\lambda) = \dfrac{d\mathcal{M}(\lambda)}{d\lambda}\dot{\lambda} - 2\mathcal{C}(\lambda) =\nonumber\\
    &\left(\hat{\mathcal{M}}-\mathcal{M}\right)\dot{\lambda} - 2\mathcal{C}(\lambda)
  \end{align}
Using parameter estimation in step (C), the structure of $\hat{\mathcal{M}}$ and $\hat{\mathcal{C}}$ are not changed. Hence, the above equation is skew-symmetric or almost skew-symmetric if and only if,
\begin{description}
  \item[a)] rate of change in decision factor is small and continues in time, or
  \item[b)] moment of inertia and mass are correctly estimated. The product of inertia and center of gravity shift estimates don't affect the stability condition.
\end{description}
When $\lambda$ is identically 0 or 1, condition (a) is granted. Either transitions $\lambda\rightarrow 0$ or $\lambda\rightarrow 1$ would occur when there is an interference of data or return to the initial model. If conditions in Remark \ref{rem:AUKF} are fulfilled the estimation error will converge to zero in finite time. Therefore, in transition, the term $\hat{\mathcal{M}}-\mathcal{M}$ has value, and nonzero $\dot{\lambda}$ will introduce an additional term in control law that may lead to instability. Accordingly, the decision factor should enter with a lag time with respect to the estimation time horizon. In this way, the temporal difference in estimation and decision-making actions would guarantee stability. The closed-loop performance is guaranteed if the robustness gain $\chi$ compensates the additional feedforward errors in the estimation of $D$ and $\tau_0$.
\qed
\end{myproof}
\begin{rem}
  Since the Lyapunov function in \eqref{eq:lyapunov} has closed level sets, in case of instability in tolerable finite time, i.e. restoring the skew-symmetric property of  $\dot{\mathcal{M}}(\lambda) - 2\mathcal{C}(\lambda)$ in small finite time, the error trajectories will return to origin after the restore. We used the qualitative term "small/tolerable finite time" because it depends on the divergence rate of flight trajectories and velocities.
\end{rem}

Eventually, for the calculation of control inputs, using the same relationship for computing optimum $\delta$ from $\tau_c$ in DAC we have,
\begin{align}\label{equ:delta_ast_lambda}
  \delta^\ast = (B(\lambda^\ast)^TB(\lambda^\ast))^{-1}B(\lambda^\ast)^T\tau_c(\lambda^\ast)
\end{align}
 The only difference would be the possibility of order reduction in effective control inputs due to control loss in case of damage. Therefore, following additional condition is imposed to guaranty non-singularity in $B(\lambda^\ast)^TB(\lambda^\ast)$,
\begin{rem}
  Assume $i$-th column of $B(\lambda^\ast)$ be denoted by $b_i$, and $\epsilon_\delta$ is a small parameter selected according to actuator gain. If $\|b_i\|\leq \epsilon_\delta$, the  $\delta_i$ and  $b_i$ will be redacted from computation of equation \eqref{equ:delta_ast_lambda}. The $\delta_i$ will keep the last value. This case should happen only when $\lambda\rightarrow 1$, if the estimation of $\hat{B}$ converges properly. Otherwise, decision factor should be decreased to give more credit to the model, and accordingly, the redacted order will return in computation.
\end{rem}


\section{SIMULATIONS}\label{sec:simulations}
Through the above steps, the DAC framework has been applied to the GTM. In this section, its performance with the following events in fault case scenario is going to be evaluated:
\begin{description}
  \item[a)] flight journey begins in trimmed condition, at $t=4$ [sec], a small disturbance is applied on the flight control surfaces to check velocity regulator performance,
  \item[b)] at $t=10$ [sec], the failure introduced in Step C happens abruptly,
  \item[c)] at $t=30$ [sec], an extra high frequency nonlinear term $5\sin(10r)\delta_r$ are added exponentially with time constant of 2 seconds to the yaw moment,
  \item[c)] at $t=50$ [sec], pilot decides to select pure MBC and then returns to DAC at $t=55$ [sec].
\end{description}

\begin{figure*}[!t]
  \centering
  \includegraphics[width=\textwidth]{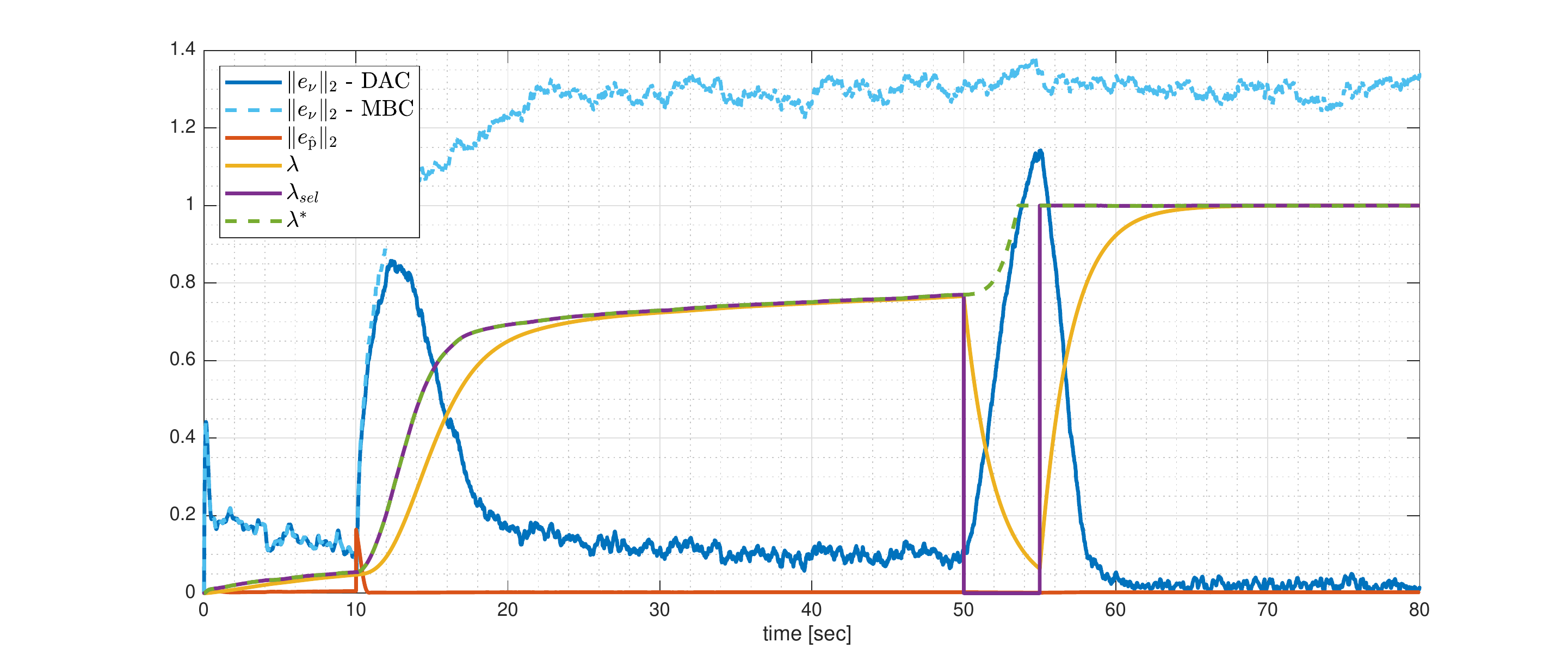}
  \caption{DAC simulation for GTM under damage and pilot decision}\label{fig:dac_per_1}
\end{figure*}

\begin{figure}[h]
  \centering
  \includegraphics[width=0.85\columnwidth]{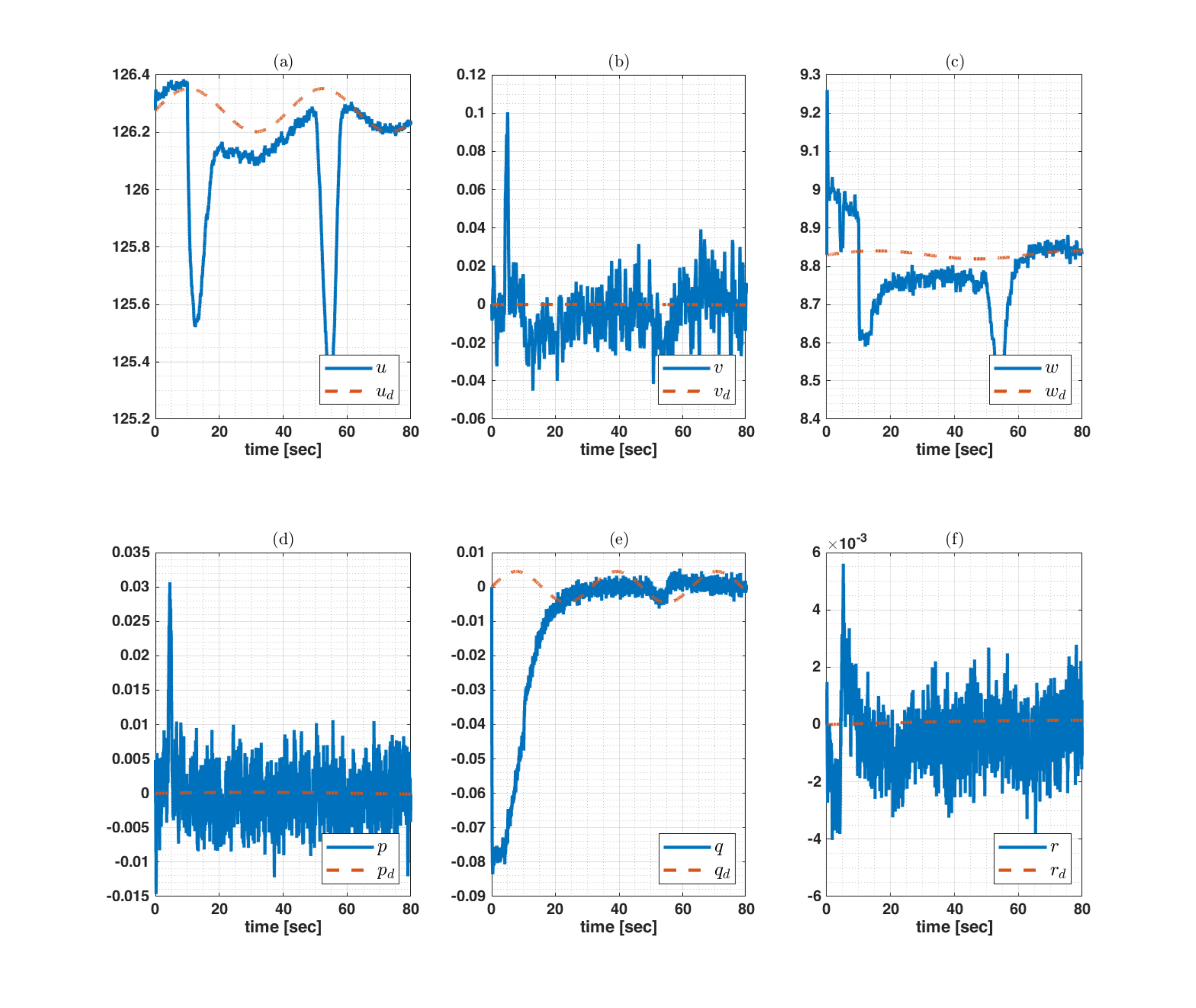}
  \caption{Velocity regulator performance with detail of state errors}\label{fig:states_tracking}
\end{figure}

\begin{figure}[h]
  \centering
  \includegraphics[width=0.7\columnwidth]{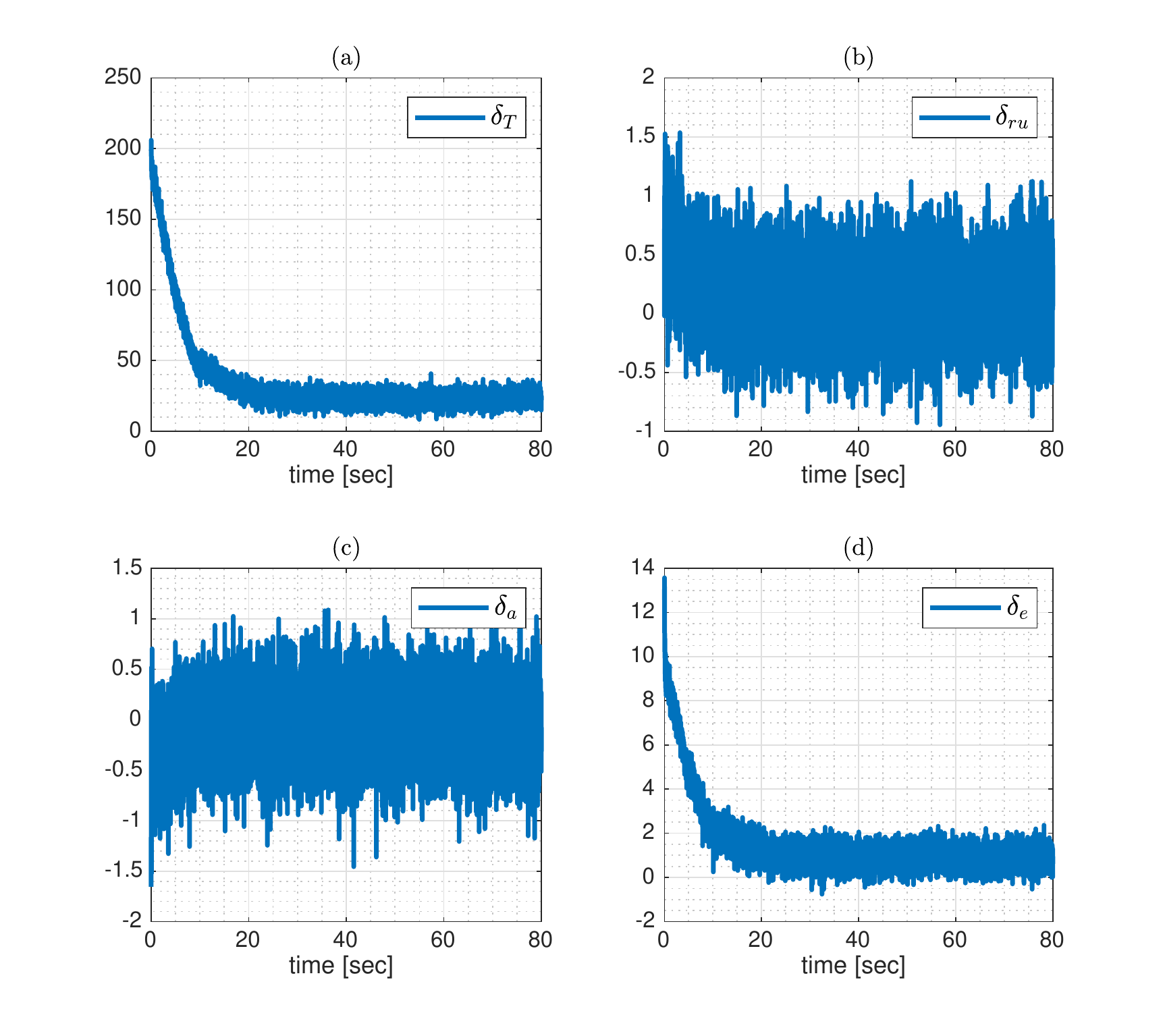}
  \caption{Control deflections}\label{fig:delta}
\end{figure}

Figure \ref{fig:dac_per_1} demonstrates simulation results of the given scenario. The effective error vanishing in DAC-enabled mode is evident, whereas the pure MBC has meaningful errors after the introduction of failure. In this plot, $\lambda$ is the actual decision factor involved in control law, and the $\lambda_{sel}$ is the selected decision factor by the pilot. The time lag between actual and optimum/selected decision factor is due to imposing intentional temporal difference in case of an error in estimations. The result implies selecting the pure MBC by the pilot will have the penalty of performance loss, however, after returning to the optimum decision with complete interference of data, i.e. $\lambda=1$, the error has vanished completely. Figure \ref{fig:states_tracking} and \ref{fig:delta} demonstrate the velocity regulator performance with details of the state errors and actuator positions respectively. Refer to \ref{rem:pe}, the sinusoidal desired values and random noise are intentionally added to ensure persistency in excitation and observability.


\begin{figure}[H]
  \centering
  \includegraphics[width=\columnwidth]{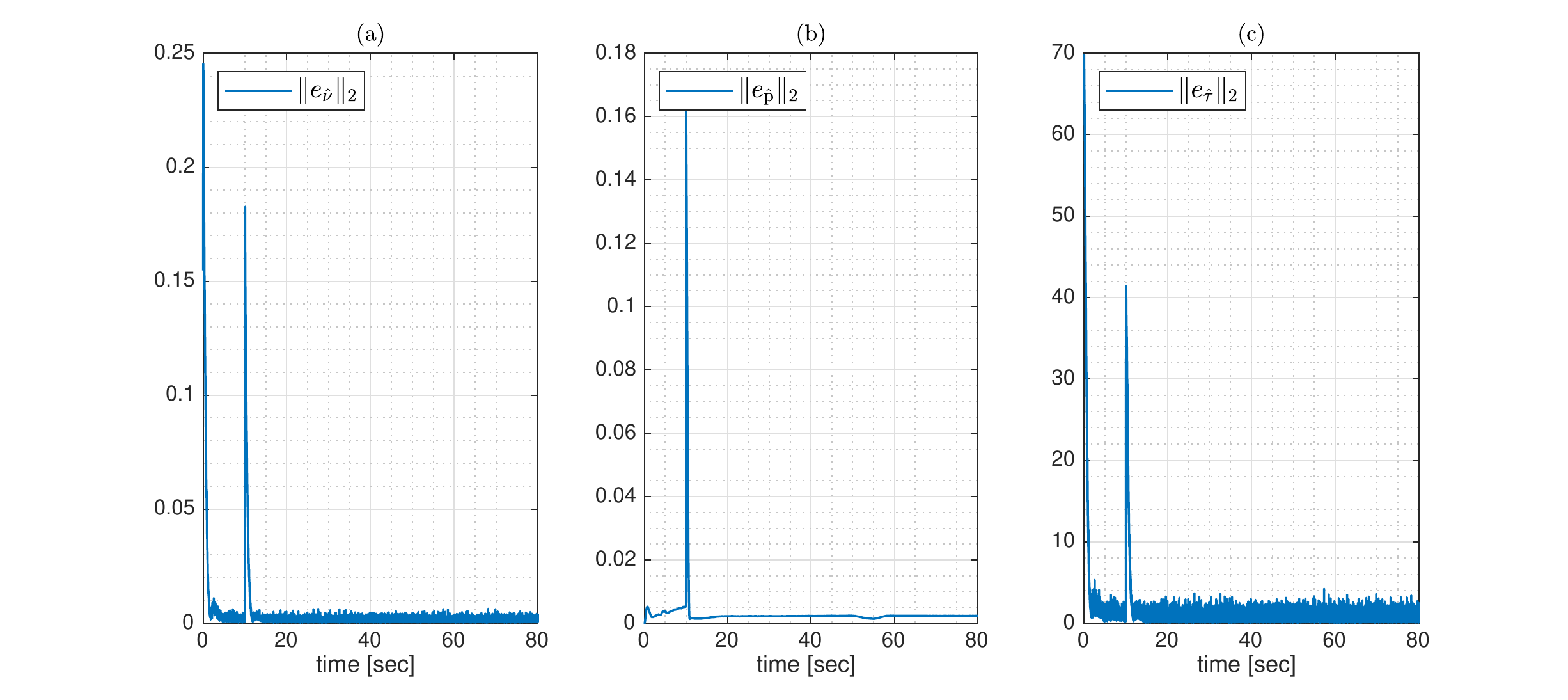}
  \caption{DUKF estimation errors}\label{fig:err_nu_p_tau}
\end{figure}

\begin{figure}[H]
  \centering
  \includegraphics[width=\columnwidth]{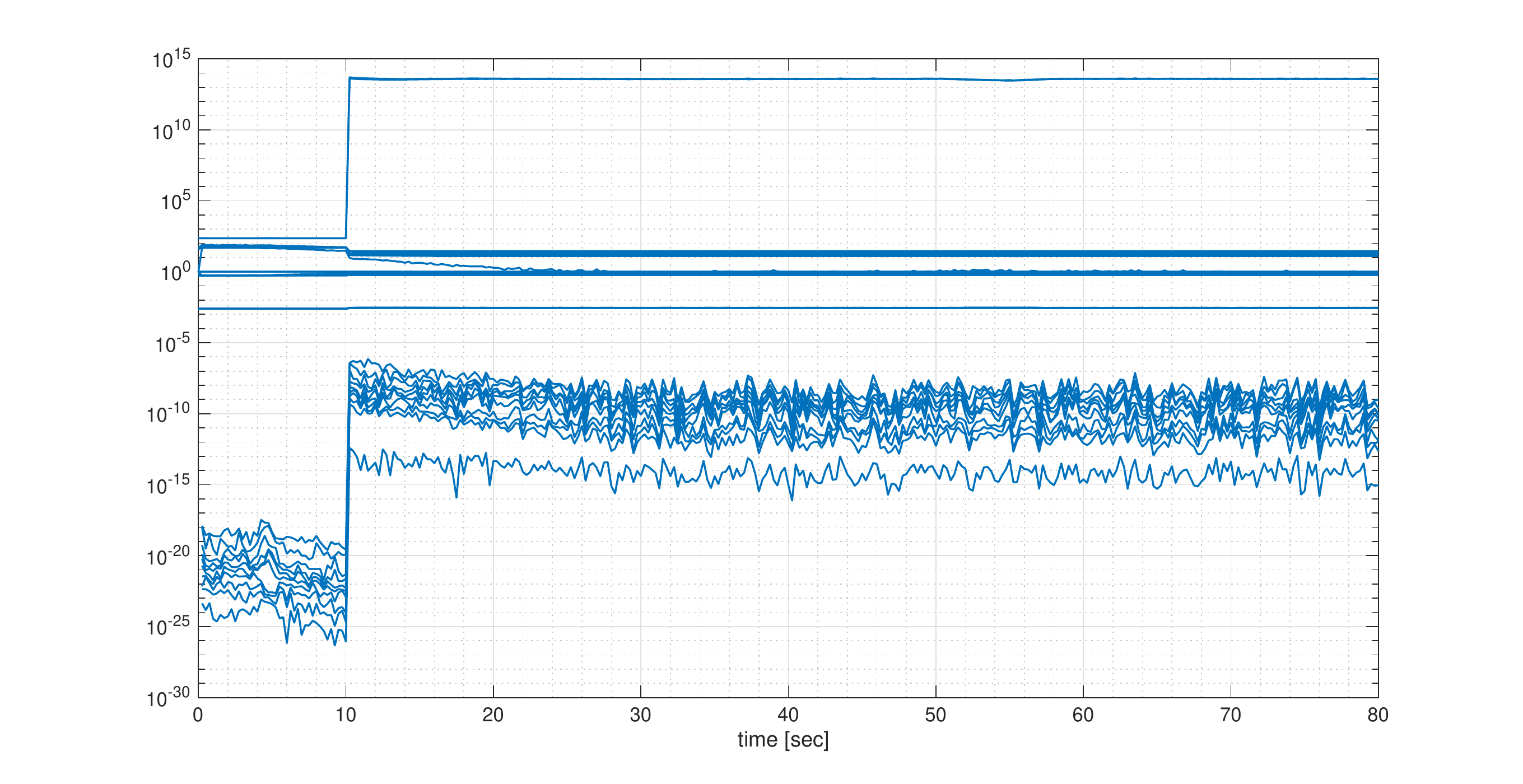}
  \caption{Singular values of the observability matrix $\mathcal{O}$}\label{fig:sigma_observability}
\end{figure}

Figure \ref{fig:err_nu_p_tau} shows an almost error-free estimation of velocities and parameters with slight error in the force-moment vector, before and after the failure. The singular values of the observability matrix are provided in Figure \ref{fig:sigma_observability}. Plot implies the observability is ascertained, however singular values of the parameters are undersized and yields weak condition of observability. It is worth mentioning when failure introduces the state and parameter variations improve the observability.

\begin{figure}[h]
  \centering
  \includegraphics[width=\columnwidth]{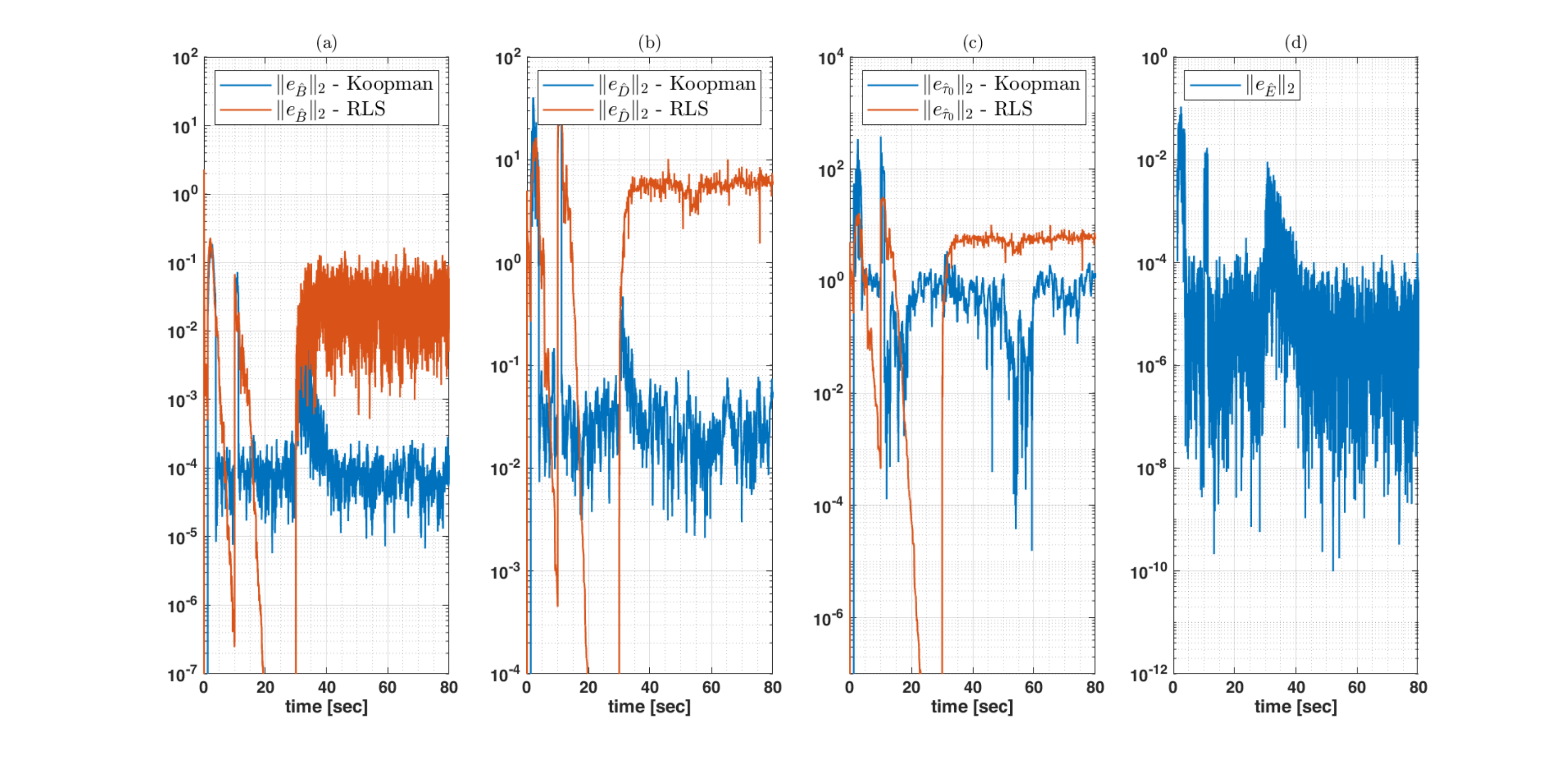}
  \caption{Comparison of the observation estimations via Koopman and RLS}\label{fig:err_koopman_vs_rls}
\end{figure}

\begin{figure}[h]
  \centering
  \includegraphics[width=0.75\columnwidth]{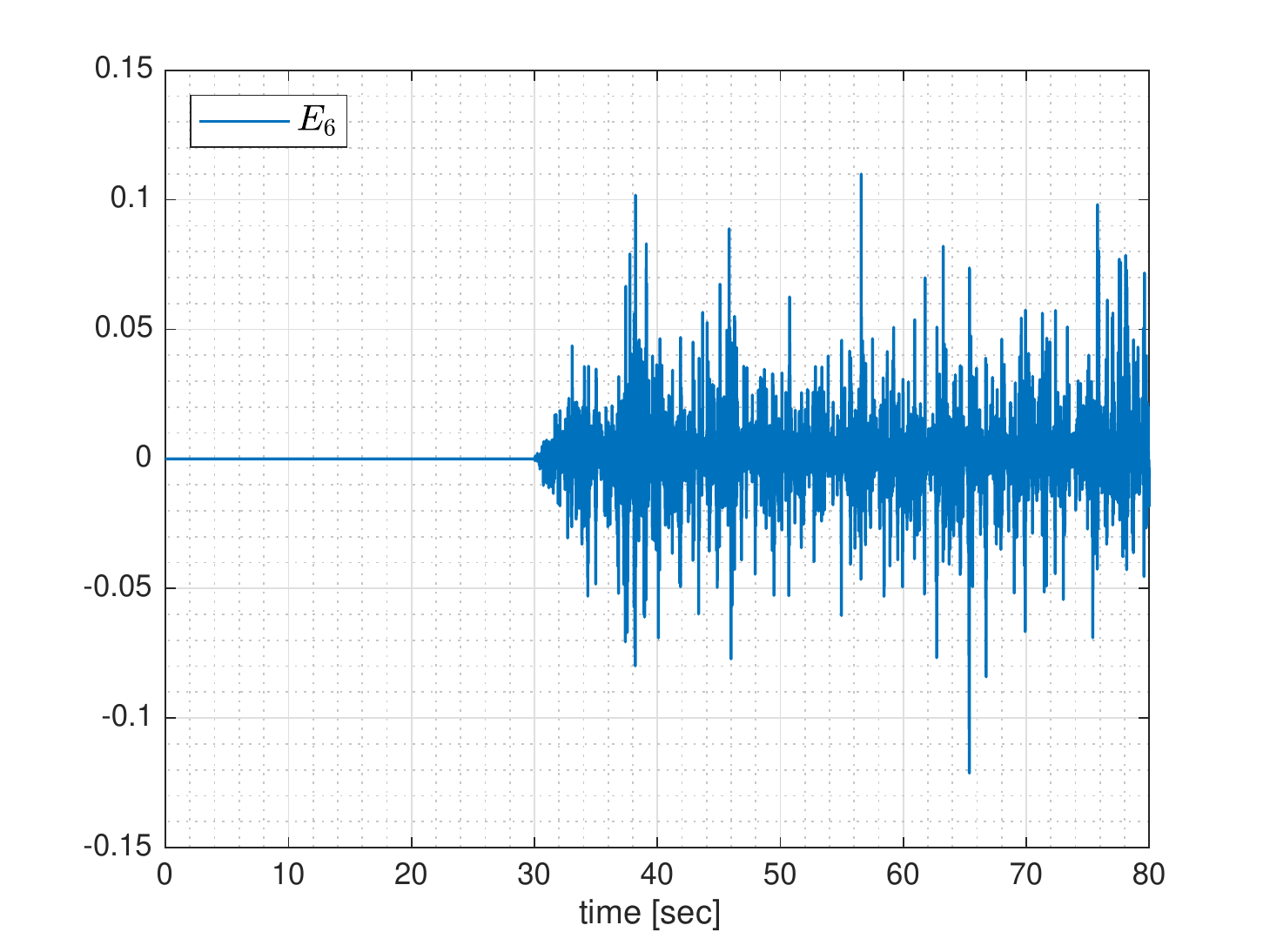}
  \caption{The high-frequency nonlinear extra term $10\sin(10r)\delta_r$ in yaw moment}\label{fig:E_term}
\end{figure}

The performance of the Koopman estimator is evaluated and compared with RLS in Figure \ref{fig:err_koopman_vs_rls}. Plots (a) to (c) demonstrate a comparative difference in the estimation of errors for $\hat{B}$, $\hat{D}$ and $\hat{\tau}_0$, after introducing a high-frequency nonlinear term in yaw moment. The amplitude of the appended term is insignificant as depicted in Figure \ref{fig:E_term}; however, the effect on the estimation result is noteworthy.  Plot (d) of Figure \ref{fig:err_koopman_vs_rls} displays estimation error of the extra term by Koopman.





\section{CONCLUSION}
As the synopsis of the study over advantage of data in the aerospace vehicle control systems, the following conclusions are realized:
\begin{itemize}
  \item using full use of available data in the future control systems is a must,
  \item employing the physical relations influence the performance of the control system which data cannot afford, particularly when dynamics are certain,
  \item the question is not using DDC or MBC. The correct question is when/where to use DDC or MBC,
  \item the flight dynamics has compatibility to be decoupled for combination use of data and model in control system analysis and design.
\end{itemize}
In this paper, the DAC framework is suggested as a comprehension of the authors regarding the above conclusions. The framework is developed exploiting the NASA GTM platform and includes a nonlinear model suite for a Lyapunov-based nonlinear velocity regulator, a dual estimation with DUKF over an observable estimation model to provide the dynamics decoupling, the Koopman estimator for identification of the linear evolution of the predefined terms over the decoupled dynamics, and eventually decision-making for the interference of data or using initial dynamical model according to the behavior of the flight. The GTM went through a series of closed-loop simulations for evaluation of the DAC performance under a failure scenario. The results show the potential and advantage of the DAC when failure happens. The stability of the closed-loop system is ascertained under specific assumptions that were full-filled in these simulations. Moreover, the estimation model is observable and the Koopman estimator is promising to map the pseudo observations into linear evolution of states, actuator positions, their derivatives and multiplications, and any feasible nonlinear combination of these variables. The obtained linear evolution is ready for updating closed-loop dynamics as the decision-making algorithm faces incapacitated behavior.

\bibliography{DAC_arxive}
\bibliographystyle{IEEEtran}






\end{document}